# Spatial-Temporal Recurrent Graph Neural Networks for Fault Diagnostics in Power Distribution Systems

Bang L. H. Nguyen, *Student Member, IEEE*, Tuyen V. Vu, *Member, IEEE*, Thai-Thanh Nguyen, *Member, IEEE*, Mayank Panwar, *Member, IEEE*, Rob Hovsapian, *Senior Member, IEEE*.

***Abstract*—Fault diagnostics are extremely important to decide proper actions toward fault isolation and system restoration. The growing integration of inverter-based distributed energy resources imposes strong influences on fault detection using traditional overcurrent relays. This paper utilizes emerging graph learning techniques to build a new temporal recurrent graph neural network models for fault diagnostics. The temporal recurrent graph neural network structures can extract the spatial-temporal features from data of voltage measurement units installed at the critical buses. From these features, fault event detection, fault type/phase classification, and fault location are performed. Compared with previous works, the proposed temporal recurrent graph neural networks provide a better generalization for fault diagnostics. Moreover, the proposed scheme retrieves the voltage signals instead of current signals so that there is no need to install relays at all lines of the distribution system. Therefore, the proposed scheme is generalizable and not limited by the number of relays installed. The effectiveness of the proposed method is comprehensively evaluated on the Potsdam microgrid and IEEE 123-node system in comparison with other neural network structures.**

## I. Introduction

PROTECTION and restoration play critical roles to enhance the resilient and reliable operation of distribution systems [1], [2]. Under the increasing integration of distributed energy resources, the protection of distribution systems becomes challenging since traditional protective relays are ineffective due to the smaller fault currents of inverter-based generators [3]. In parallel with passive relays, fault diagnostics using measurement data targets provide system operators with fault types and locations to timely isolate faults and restore normal operations [4].

Fault diagnostics include fault event detection, fault type/phase classification, and fault location. There are many fault diagnostics schemes analyzing data from digital relays or micro phasor measurement units (μPMU) proposed in literature [3]–[18]. Loosely, these schemes can be classified into model-based and data-driven based techniques.

Model-based methods focus on finding the evaluation metrics that are consistent in accordance with the proposed fault models. In [19], the pre-fault negative sequence and positive sequence current are compared for detection. However, the performance of this method is significantly affected by the fault current amplitudes. This requires readjustment and prior information about all possible microgrid configurations to determine an appropriate threshold. A transient monitoring function to detect fault is proposed in [20] by summing the residuals between estimated and measured current components over one cycle. Although this differential-based method does not rely on the magnitude of the fault current, the unbalanced loads and generation transients can cause a false alarm. In [21], the mathematical morphology and recursive least square are employed. The Teager-Kaiser energy operator is proposed in [22] to detect and classify faults. These methods are strongly dependent on the configuration of the distribution system and cannot find the fault location.

Data-driven methods focus on mining measurement data to diagnose faults. The decision tree [23] and random forest [24], which are popular statistical classifiers, have been applied in fault detection. In [25], four machine-learning classifiers i.e., decision tree, K-nearest neighbor, support vector machine, and Naïve Bayes are implemented and compared for fault diagnostics. The discrete wavelet transform is frequently employed as a feature extraction technique [26] prior to the classification process. Advanced machine learning techniques are also adopted recently i.e., the maximal overlap discrete wavelet transform and extreme gradient boost algorithm in [27], Taguchi-based artificial neural networks in [28], and gated-recurrent-unit deep neural networks in [29] and are achieved very high accuracy. However, in these works, fault detection, classification, and location are performed based on the current measurements from the fault line. There is a research gap in fault diagnostic in distribution systems with limited data where the faults may occur in lines without measurement devices.

Fault diagnostics using PMU data have been investigated in several papers. In [30], [31], the fault location is determined based on the discrepancy of the nodal voltages calculated based on μPMU data and pseudo-measurements. The accuracy

B. L. H. Nguyen and T. V, Vu are with Clarkson University, Potsdam, NY, USA (e-mail: nguyenbl@clarkson.edu, dassc@clarkson.edu, hanq@clarkson.edu, tvu@clarkson.edu). T.-T. Nguyen is with New York Power Authority (e-mail: thaithanh.nguyen@nypa.gov). M. Panwar and R. Hovsapian are with National Renewable Energy Laboratory. (e-mail: mayank.panwar@nrel.gov, rob.hovsapian@nrel.gov)

of this method depends on the load model and the reliability of pseudo measurements. With a larger scope, data from two µPMUs are analyzed to locate and classify events in the distribution grid using SVM, k-NN, and DT algorithms [32]. However, the investigated system is radial with small nodes, and the number of events is small. The faulted line location using µPMU data via convolutional neural networks is proposed in [33], [34]. The semi-supervised learning is performed on µPMU data to detect and locate high-impedance faults [35]. None of the mentioned works demonstrated fault diagnostics on mesh-topology networks and their scheme lack fault type/phase classification.

This paper proposes a unified fault diagnostic scheme including detection, classification, and location based on voltage measurement data, which can be collected from µPMU, advanced metering infrastructure (AMI), and consumer-side smart meters. The proposed scheme leverages transfer learning and fine-tuning with the combination of recurrent neural networks (RNN) and graph neural networks for the diagnostic models. Although there are existing fault detection schemes using graph neural networks (GNN) or graph convolutional networks (GCN), those works contain limits or focus on different objects as follows. [34] only focus on the fault location and lack of comprehensive analysis of the results. [36] applies the GNN for fault diagnosis of transformers. Moreover, none of the existing works have considered the temporal correlation in graph learning on time-series data of fault diagnostic problems.

The unique contributions are outlined as follows

- The combination of RNN and GNN structures is proposed for fault diagnostics with voltage measurement data as inputs.
- Both spatial and temporal correlations in the graph-based time-series data are intrinsically considered by the temporal recurrent graph neural network (R-GNN).
- The proposed fault diagnostic scheme can detect fault events, classify the fault type and phase, and identify the fault location.
- The transfer learning and fine-tuning approaches are implemented for multiple learning tasks in the proposed fault diagnostic scheme.
- Comprehensive case studies and comparisons with other machine learning techniques and NN structures such as general artificial NN (ANN), RNN, convolutional NN (CNN), and GCN are also provided.

Notably, the proposed deep NN structure is capable of incorporating current measurements as edge-feature inputs for fault detection; however, this is not investigated in this paper but future work. The remaining parts are organized as follows. In Section II, the Potsdam microgrids and IEEE 123-node feeder systems under investigation are described. Section III introduces the deep graph neural network model considering spatial-temporal µPMU data. Section IV describes the data collection, pre-processing, and series of case studies to demonstrate the effectiveness of the proposed scheme. The results are compared and discussed in Section V. Section IV concludes the paper.

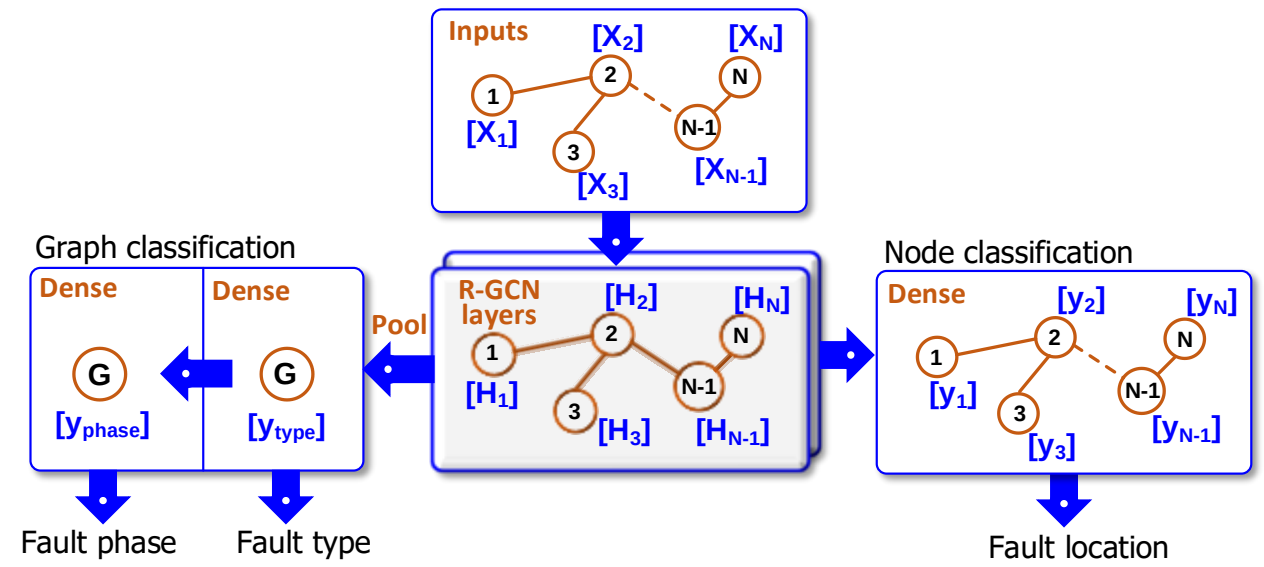

Fig. 1. Block diagram of the proposed fault diagnostic scheme.

## II. Proposed Fault Diagnostics Scheme Using Temporal Recurrent Graph Neural Networks

### A. Preliminaries

The distribution network is defined as an undirected graph $\mathcal{G} = (\mathcal{V}, \mathcal{E}, \mathcal{A})$, where $\mathcal{V}$ denotes the set of vertices, $|\mathcal{V}| = N$, each vertex in the graph represents a node (bus) in the distribution network, $X = \{X_1, X_2, \dots X_N\}$ is the tuple of node features, $\mathcal{E}$ denotes the set of edges, $|\mathcal{E}| = M$, each edge represents a line (branch) connecting two buses, $E = \{E_1, E_2, \dots E_M\}$ is the tuple of edge feature, and $\mathcal{A} \in \mathbb{R}^{N\times N}$ denotes the adjacency matrix of the distribution network. The input data for graph learning are the node features $X_{i=1\dots N}$, and the edge features $E_{i=1\dots M}$. Some applications also contain the attributes for each graph data $(u)$ [37].

The learning goal is to generalize the mapping model between the inputs of node and/or edge attributes and the outputs. The outputs of graph learning can be the classification or regression task at node or graph levels. The input-output model $\mathcal{F}(\cdot)$ of the GCN can be expressed as

$$\hat{y} = \mathcal{F}(\mathcal{G}, \mathcal{V}, \mathcal{E}, \mathcal{W}), \quad (1)$$

where $\mathcal{W}$ is the trainable weights, $\hat{y}$ is the inferred output. The trainable weights are updated iteratively via backpropagation over minimizing the loss function $\mathcal{L}(\hat{y}, y)$, where $y$ is the output labels. The loss function can be a mean squared error (MSE) or mean absolute error (MAE) in a regression problem or cross-entropy in a classification problem [38].

The main difference between the traditional and graph neural networks structures is that graph learning includes the graph structure via the adjacency matrix $\mathcal{A}$ of the undirected graph $\mathcal{G} = (\mathcal{V}, \mathcal{E}, \mathcal{A})$. In case of that, the set of edges $\mathcal{E}$ and the adjacency matrix $\mathcal{A}$ do not change, we have a static graph. Otherwise, there is a dynamic graph [39].

### B. Fault Diagnostic Scheme via Recurrent-Graph Neural Networks

This paper focuses on fault diagnosis via graph neural network models by voltage measurements. Herein, only the bus voltage measurements are considered as input node features for consistency. The incorporation of current measurement as the edge features would be extended in future work. Each bus have three-phase voltages $V_{a,k}$, $V_{b,k}$, $V_{c,k}$ in time series, where $k$ is the time index. Therefore, the node feature in node $i$ is shown in the form of

$$X_i = \begin{bmatrix} V_{a,1} & V_{a,2} & \cdots & V_{a,K} \\ V_{b,1} & V_{b,2} & \cdots & V_{b,K} \\ V_{c,1} & V_{c,2} & \cdots & V_{c,K} \end{bmatrix}^T, \quad (2)$$

where $K$ is the length of the evaluation period. The learning performance is investigated under different values of $K$. It is worth noting that the nodal admittance matrix $(Y)$ of the distribution network can be the weighted adjacency matrix. The graph data $\mathcal{G} = (\mathcal{V}: \{X_1, X_2, \dots X_N\}, \mathcal{A})$ includes the voltage measurement of nodes in the distribution system and the adjacency matrix $\mathcal{A}$ representing the connection of the graph.

Outputs $\hat{y}$ of the GCN models $\mathcal{F}(\cdot)$ are the fault categories and fault location, where the graph classification task is for classifying the fault categories and the node classification is for determining the fault location. The fault categories have two labels: fault types and fault phases. The fault types are classified into six types included **1)** no-fault (**NF**), **2)** single-phase-to-ground (**LG**), **3)** two-phase (**LL**), **4)** two-phase-to-ground (**LLG**), **5)** three-phase (**3L**), and **6)** three-phase-to-ground (**3LG**). Therefore, $y_{type} \in \mathbb{B}^{1\times 6}$ with the $i$-th element of $y_{type}$: $y_{type}[i] = 1$ indicates the $i$-th fault category occurred while all other $y_{type}[i] = 0$. The fault phases are determined by $y_{phase} \in \mathbb{B}^{1\times 3}$, where $y_{phase}[i] = 1$ indicating the fault occurs in phase $A, B, C$, or $AB, BC, CA$ when the fault types are asymmetrical i.e. LG, LL, and LLG, respectively. The fault location is indicated by $y_i = 1$, where $i = 1, 2, 3, \dots N$ if the fault occurs in the $i$-th bus, otherwise $y_i = 0$. The fault location detection is performed at node-level classification.

The block diagram of the proposed fault diagnostic scheme using recurrent graph learning is shown in Fig. 1. From the node features $\{X_1, X_2, \dots X_N\}$, the hidden features $\{H_1, H_2, \dots H_N\}$ are extracted through the R-GCN layers. From these distinct hidden features, on one hand, the fault location outputs $y_i$ can be captured using the dense layers in a node classification task. On the other hand, the pooling operation is performed to achieve the unified graph features, then via the dense layers, the fault type and fault phase is determined. The fault type is detected first, and in cases of asymmetrical faults detected, the fault phase is identified thereafter. The R-GCN layers are designed in detail in the next section *II. C*. The transfer learning and fine-tuning techniques are described in section *II. D* is applied to reduce the training time for multi-tasks in this fault diagnostic scheme.

## *C. Temporal Recurrent-GCN layers*

Existing works adopted the gated recurrent unit (GRU) or graph convolutional network (GCN) structures for fault diagnostic in the distribution system in [29] and [34], respectively. However, these structures can only extract either the temporal or spatial dependencies. This paper implements the temporal recurrent GCN layers of the graph-learning-based models for fault diagnosis. Temporal R-GCN layers can capture both temporal and spatial correlation in the input data. The fault diagnostic models are represented by a classification function $\mathcal{F}(\cdot)$ mapping the input time sires $(X_i^1, X_i^2, \dots X_i^K)$ over the graph $\mathcal{G}$ to the fault labels as follows.

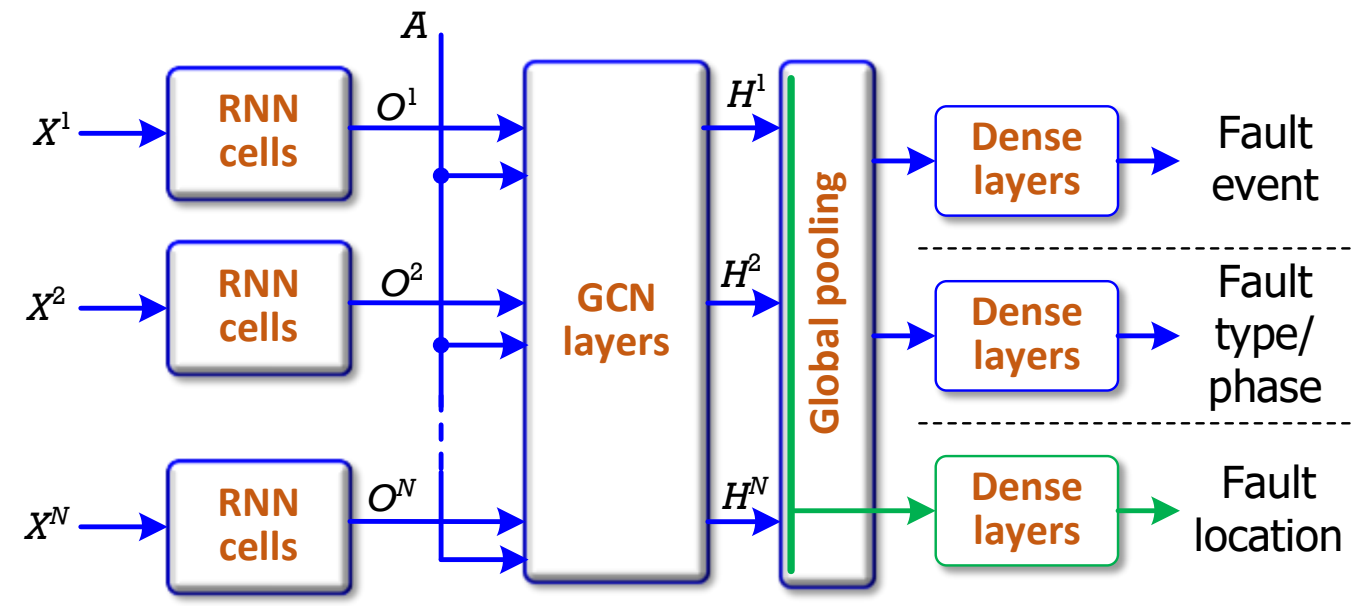

Fig. 2. Proposed temporal R-GCN structure for fault diagnostic.

$$\hat{y} = \mathcal{F}(\mathcal{V}: \{X_1, X_2, \dots X_N\}, \mathcal{A}), \quad (3)$$

Where the node feature $X_1 = \{X_1^1, X_1^2, \dots X_1^K\}$ and so on with the under script denoting the node index and the superscript denoting the time index. The structure of the graph $\mathcal{G}$ is reflected through the adjacency matrix $\mathcal{A} \in \mathbb{R}^{N\times N}$.

The proposed temporal R-GCN framework for fault diagnosis is illustrated in Fig. 2. The proposed R-GCN structure includes RNN cells and GCN layers for feature extraction. Firstly, the RNN cells are employed to extract the temporal feature from the voltage in the time series of each node. Thereafter, the GCN layers are used to identify the spatial correlation between the bus voltages over the distribution system. The global pooling operation concentrates all hidden features from nodes and finally, the dense layers are trained to classify the fault type and fault phase. The fault location is performed based on all hidden features from all the nodes. The formulation of GCN and RNN layers is presented as follows.

*Graph Convolutional Network Layers*:

The node feature at each time index is processed by the GCN layers [40], which can be expressed as

$$H_{(l+1)}^i = \sigma\left(\widetilde{D}^{-\frac{1}{2}}\tilde{A}\widetilde{D}^{-\frac{1}{2}}H_{(l)}^i W_{(l)}\right), \quad (4)$$

where $\tilde{A} = A + I_N$ is the adjacent matrix with self-connection, $I_N$ is the identity matrix, $\widetilde{D}$ is the agree to matrix from $\tilde{A}$ with $\widetilde{D}_{ii} = \sum_i \tilde{A}_{ij}$ and $\widetilde{D}_{ij} = 0$, $U_{(l)}^i$ is the output of layer $l$, $H_{(0)}^i = \mathcal{V}^i$, $W_{(l)}$ is the weight matrix of layer $l$, $\sigma(\cdot)$ is a nonlinear activation function. This graph propagation formula can be derived as a first-order approximation of localized spectral filers [37].

Outputs of GCN layers at each time index are the inputs of a recurrent neural network (RNN), where the RNN cells can be GRU or long-short-term memory (LSTM) [41]. GRU structure is simpler than LSTM, thus it is computationally more efficient. However, LSTM can remember longer sequences and achieve better performance in temporal long-distance tasks [42].

*Long-Short-Term Memory Cell*:

One LSTM cell computes for each time step the hidden state $\boldsymbol{h_t}$ and the cell state $\boldsymbol{c_t}$ from the input $\boldsymbol{x_t}$ , the previous hidden state $\boldsymbol{h_{t-1}}$, and the previous cell state $\boldsymbol{c_{t-1}}$. In each LSTM, there are intermediate states of the forget gate $\boldsymbol{f_t}$, the cell candidate $\boldsymbol{g_t}$, the input gate $\boldsymbol{i_t}$, and the output gate $\boldsymbol{o_t}$. The relationship between these state variables is expressed as follows.

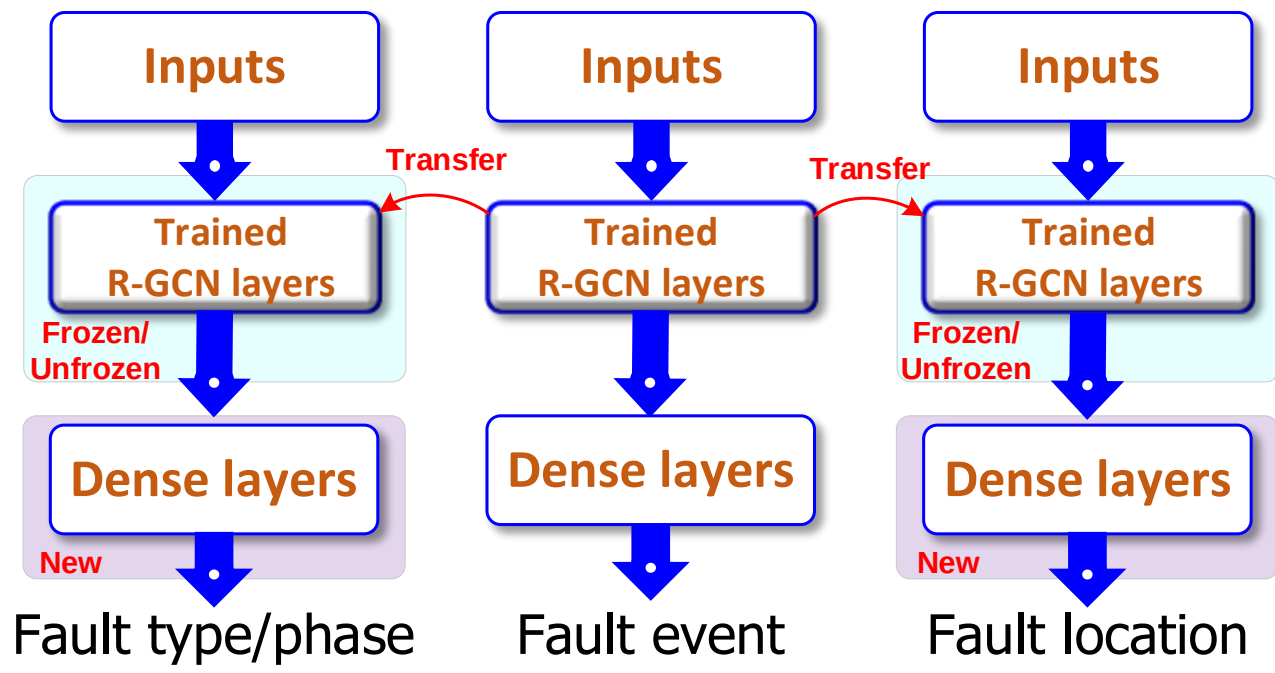

Fig. 3. Block diagram of layer transfer for multi-tasking fault diagnostic.

$$f_t = \sigma_g\left(W_f x_t + R_f h_{t-1} + b_f\right), \tag{5}$$

$$g_t = \sigma_c\left(W_g x_t + R_g h_{t-1} + b_g\right), \tag{6}$$

$$i_t = \sigma_g(W_i x_t + R_i h_{t-1} + b_i), \tag{7}$$

$$o_t = \sigma_g(W_o x_t + R_o h_{t-1} + b_o). \tag{8}$$

The matrices $W_f$, $W_g$, $W_i$, $W_o$, $R_i$, $R_g$, $R_o$, $R_i$ and the biased vectors $b_f$, $b_g$, $b_i$, $b_o$ are the trainable weights. The gate activation functions $\sigma_g(\cdot)$ is sigmoid, and $\sigma_c(\cdot)$ is tanh. The cell state and hidden state are computed as

$$c_t = f_t \circ c_{t-1} + g_t \circ i_t, \tag{9}$$

$$h_t = f_t \circ \sigma_c(c_{t-1}), \tag{10}$$

where $\circ$ denotes the element-wise product.

*Gated Recurrent Unit Cell*:

The GRU cell contains only the reset gate $\boldsymbol{r_t}$, the updated gate $\boldsymbol{z_t}$ expressed as follows.

$$r_t = \sigma_r(W_r x_t + R_r h_{t-1} + b_r), \tag{11}$$

$$z_t = \sigma_z(W_z x_t + R_z h_{t-1} + b_z), \tag{12}$$

Then, the candidate hidden state $\boldsymbol{\tilde{h}_t}$ and hidden state $\boldsymbol{h_t}$ can be computed

$$\tilde{h}_t = \tanh(W_h x_t + W_{rh}(r_t \circ h_{t-1}) + b_h), \tag{13}$$

$$h_t = z_t \circ h_{t-1} + (1 - z_t) \circ \tilde{h}_t, \tag{14}$$

The matrices $W_r$, $W_z$, $W_h$, $W_{rh}$, $R_r$, $R_z$ and the biased vectors $b_r$, $b_z$, $b_h$ are the trainable weights. $\sigma_r(\cdot)$ and $\sigma_z(\cdot)$ are the activation functions.

The hidden states of the last layers of RNN cells are collected and flattened. Thereafter, dense layers are employed to calculate the outputs. Notably, there are three dense layer blocks for each fault diagnostic task. Therefore, we have 3 different models for each task. The formulation of the dense layers is given as follows:

$$\hat{y} = \sigma_d(W_d[h_1, h_2, \dots h_L] + b_d), \tag{15}$$

where $W_d$, $b_d$ are trainable weights, $[h_1, h_2, \dots h_L]$ represents the flattened matrix of all hidden states collected from RNN cells, $\sigma_d(\cdot)$ is an activation function.

GRU has fewer trainable parameters and does not have internal memory, it is trained faster with less memory used. The LSTM has more gates and processes internal memory, it is more accurate on a large dataset. Since our dataset is quite large, in this paper, we only employ the LSTM.

### D. *Transfer Learning for Multi-tasking Fault Diagnostic*

As can be seen in Figs.1 and 2, fault diagnostic includes three different tasks: fault event detection, fault type/phase classification, and fault location identification. Traditionally, for each task, a standalone deep NN model is trained independently [29]. This approach is straightforward but inefficient. The training, implementing, and interfering processes are triple. Other approaches can improve training efficiency, reduce overfitting, and speed up the training process [43], [44].

Fig. 3 illustrates the transfer learning techniques for multi-tasking fault diagnostics. Therein, the R-GCN model of fault-event classification is trained in advance. Then, the trained R-GCN layers in this model, which are responsible for node feature extraction, are transferred to the new R-GCN models of fault-type/phase identification. Two new RGNN models consist of the trained RGNN layers and the new additional dense layers. One can freeze the weights of trained R-GNN layers in the training process of the new models so that only the weights of dense layers are trained for new tasks. However, the trained RGNN layers may be overfitting to the trained task and cause negative transfer effects, which makes the training process for later tasks harder [45]. To alleviate this problem, we trained the fault event classification to only 95% accuracy and then froze the R-GCN layers to transfer them to train other tasks. First, we keep these R-GCN layers frozen and train appropriate dense layers for fault type/phase classification and fault location tasks. Thereafter, we unfreeze the transferred R-GNN layers and do fine-tune them by using a small learning rate of 0.001 to train all layers [44]. Therefore, we adopt these two transfer learning techniques including 1) layer-transferring and 2) fine-tuning to reduce the training efforts.

## III. GNN-Based Fault Diagnostic Implementation

### A. *Investigated Distribution Systems*

The comprehensive case studies in this paper focus on the Potsdam microgrid [46] and IEEE 123-node feeder as shown in Figs. 4 and 5. The Potsdam microgrid consists of 5 inverter-based generators (IBG) operating in the islanded mode under a primary droop control strategy [47] and secondary PI controller for frequency and average voltage regulation [48]. The line-line voltage level is 13.2 kV at 60 Hz. The parameters of loads and IBGs are set following parameters in [49]. The voltage measurements are recorded in buses marked with a blue square. Thereafter, we reduce the number of voltage measurement inputs to verify the performance of the trained fault diagnostic models. The bus voltages are sampled at a 1 kHz rate at the corresponding voltage measurement in devices via instrument transformers. The entire microgrid system is simulated in real-time using Opal-RT. The operational data of load changes and faults under different scenarios are collected for training and testing processes in the proposed fault diagnostic scheme using deep graph neural networks. The graph structure for graph data in Potsdam microgrid is built based on all 13 buses.

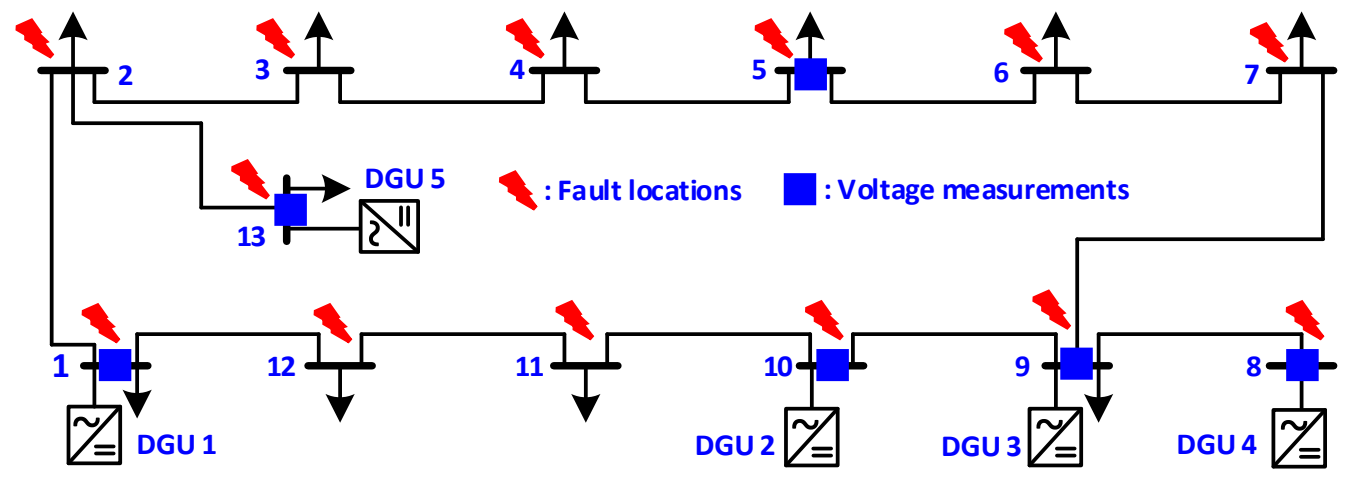

Fig. 4. 13-bus Potsdam microgrid system diagram with fault locations and voltage measurements on buses 1, 5, 8, 9, 10, and 13.

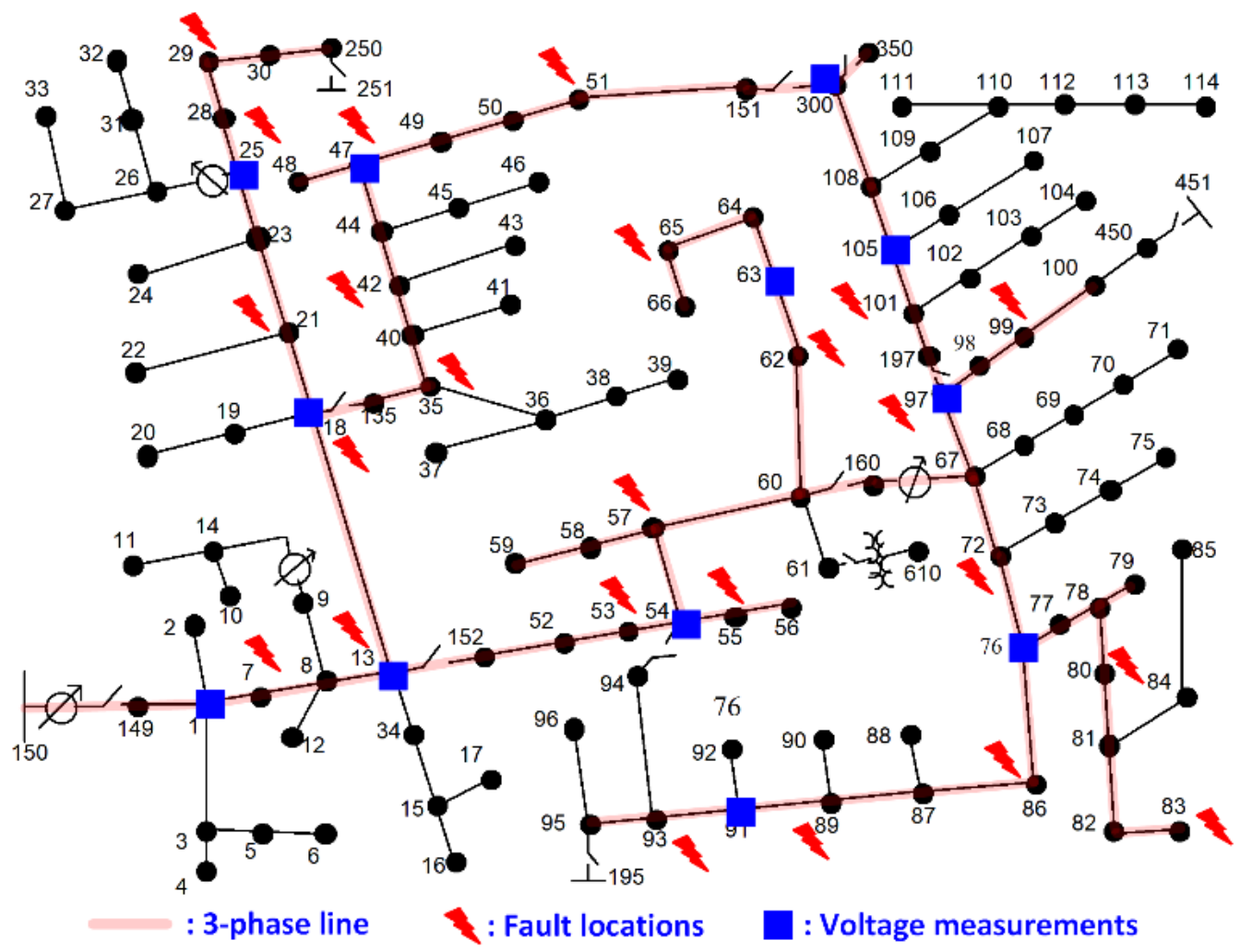

Fig. 5. IEEE 123-node feeder with fault locations and voltage measurement in 3-phase lines at buses 1, 13, 18, 25, 47, 54, 63, 76, 91, 97, 105 and 300.

To prove the scalability of the proposed scheme, the IEEE 123-node feeder [50] is also constructed in the Opal-RT real-time simulator. Fault locations are placed at buses in three-phase lines scattered over the system as shown in Fig. 5. The voltage measurements are also recorded only on buses marked with a blue square. The graph structure for graph data in IEEE 123-node feeder is built based on the connection of only 46 main buses (1, 7, 8, 13, 18, 21, 23, 25, 28, 29, 30, 35, 40, 42, 44, 47, 49, 50, 51, 52, 53, 54, 57, 60, 62, 63, 66, 67, 72, 76, 78, 81, 82, 83, 86, 87, 89, 91, 93, 95, 97, 99, 101, 105, 108, 300). Notably, this IEEE 123 node-feeder system is slightly

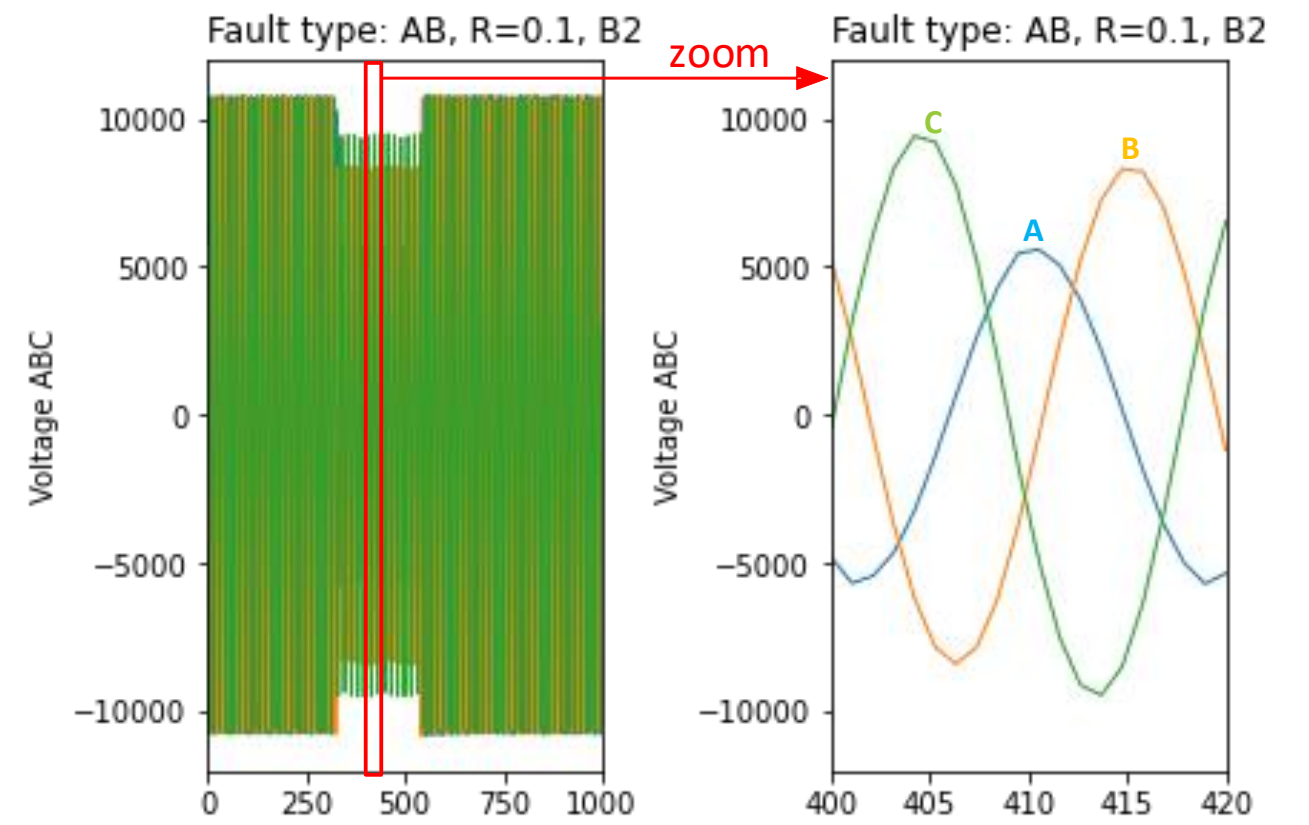

Fig. 6. PMU voltage waveform in phases A, B, C at bus 1 with AB fault and fault resistance 0.1 Ω occurs at bus 2 in Potsdam microgrid.

TABLE I. POTSDAM MICROGRID DATASET

| Parameters | Configuration | Count |
|---|---|---|
| **Fault type** | AG, BG, CG, AB, BC, CA, ABG, BCG, CAG, ABC, ABCG | 11 |
| **Fault resistance** | 0.1, 1 , 10 (Ω) | 3 |
| **Fault location** | **Buses**: 1, 2, 3, 4, 5, 6, 7, 8, 9, 10, 11, 12, 13. | 13 |
| **Load scenario** | randomly | 150 |
| **Total fault cases:** 64,350 \| **Train:** 55,770 \| **Test:** 8,580 | | |
| **Total load change cases:** 10, 000 \| **Train:** 8,580 \| **Test:** 1,420 | | |
| **Train-set**: 64,350 samples \| **Test-set**: 10,000 samples | | |

TABLE II. 123-NODE FEEDER DATASET

| Parameters | Configuration | Count |
|---|---|---|
| **Fault type** | AG, BG, CG, AB, BC, CA, ABG, BCG, CAG, ABC, ABCG | 11 |
| **Fault resistance** | 0.1, 1 , 10 (Ω) | 3 |
| **Fault location** | **Buses**: 7,13, 18, 21, 25, 29, 35, 42, 47, 51, 53, 55, 57, 62, 65, 72, 80, 83, 86, 89, 93, 97, 99, 101. | 25 |
| **Load scenario** | randomly | 50 |
| **Total fault cases:** 41,250 \| **Train:** 33,000 \| **Test:** 8,250 | | |
| **Total load change cases:** 10, 000 \| **Train:** 8,250 \| **Test:** 1,750 | | |
| **Train-set**: 41,250 samples \| **Test-set**: 10,000 samples | | |

unbalanced. However, since the investigated fault resistances are not so high, the voltage drops are still significant to distinguish the faults. High-impedance faults under such an unbalanced system will be investigated comprehensively in future works.

## *B. Temporal Graph Dataset*

The temporal graph dataset is constructed by the ordered set of graph, node feature matrix, and label vector tuples [39] $\mathcal{D} = \{(\mathcal{G}^1, X^1, y^1), (\mathcal{G}^2, X^2, y^2), \ldots (\mathcal{G}^I, X^I, y^I)\}$, where the vertex sets is unchanged $\mathcal{V}^i = \mathcal{V}, \forall i \in \{1, \ldots, I\}$, $i$ is the graph data index . The node feature matrices $X^i \in \mathbb{R}^{N \times d \times K}$ have 3 dimensions as follows: the number of nodes $|\mathcal{V}| = N$, the number of features in each node $d$, and the time interval $K$. The label vector includes 3 labels of the distribution network

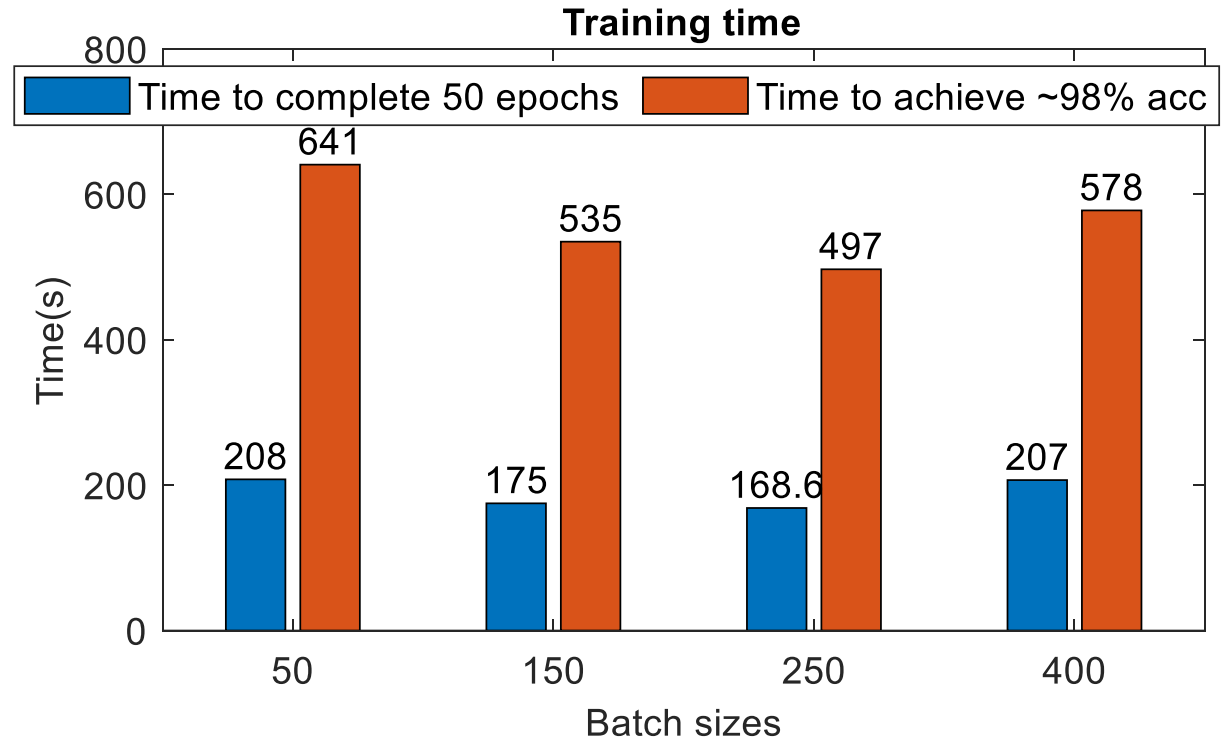

Fig. 7. Comparison of training time to complete 50 epochs and time to achieve about 98% accuracy of fault event detection in Potsdam microgrid under different batch sizes.

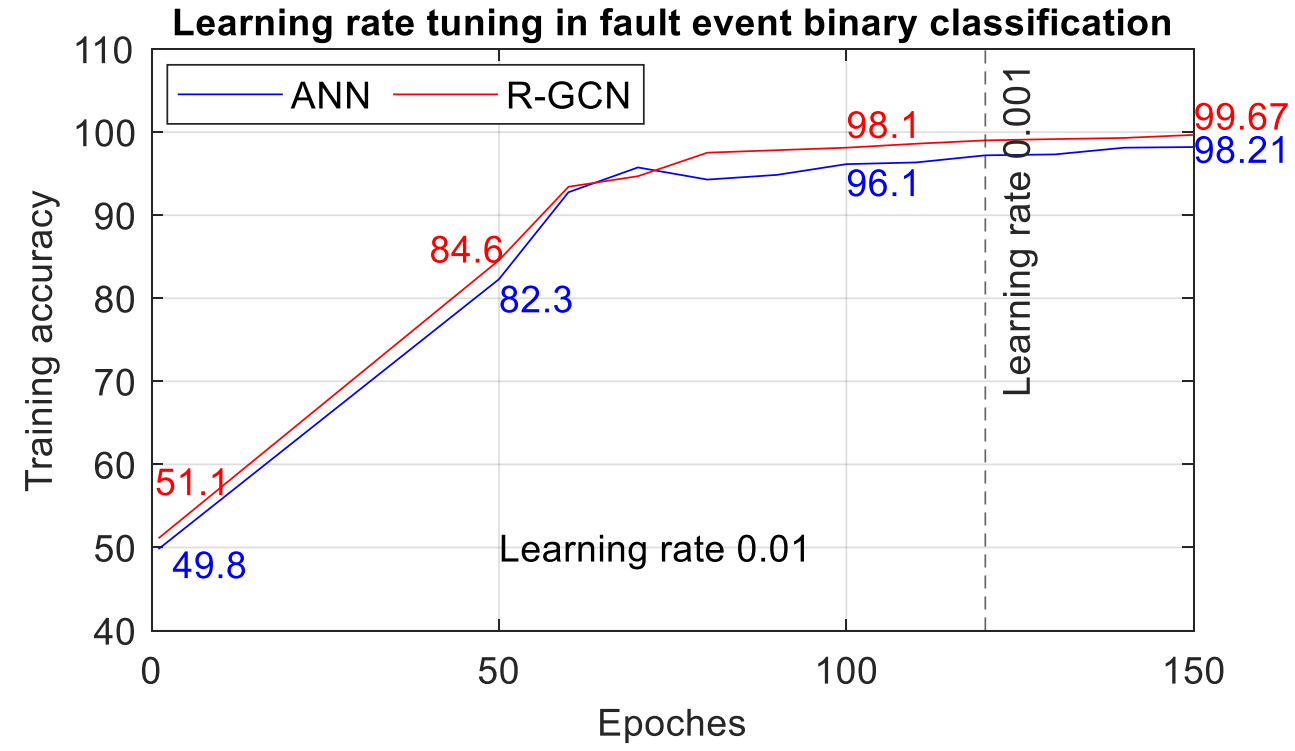

Fig. 8. The training accuracy curves with ANN and R-GCN structures.

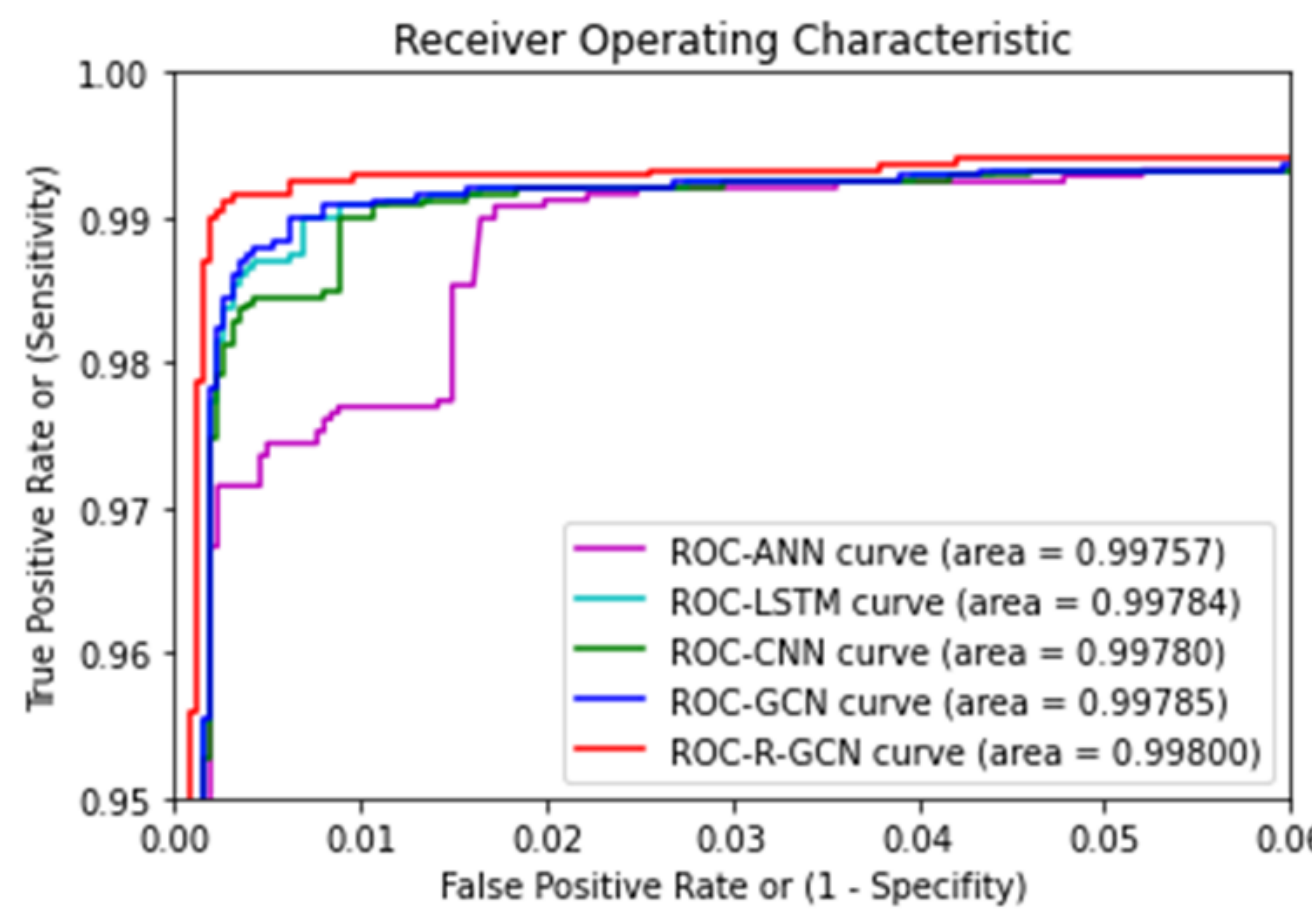

Fig. 9. The receiver operating characteristic (ROC) curves and area under curve (AUC) of ANN, LSTM, CNN, GCN, R-GCN fault event detection.

graph over the time interval $K$, $y^i = \{y_{type}, y_{phase}, y_{loc}\}$, where $y_{loc}$ is the node index where the fault occurs. The node feature matrix $X^i = \{X_1, X_2, \dots X_N\}$ contains the bus voltages of all measured buses. In the bus without voltage measured, the node features are filled with zeros. The diagnostic performances are compared between these three tasks under different numbers of voltage measurements. Fig. 6 shows the time-series voltage captured from Opal-RT real-time simulator in a 1-second time window with a fault occurring in between. There is a total of 64,350 data windows captured for 11 fault types (AG, BG, CG, AB, BC, CA, ABG, BCG, CAG, ABC, and ABCG), 3 fault resistance (0.1, 1, and 10 Ω) occurred at 13 buses of Potsdam microgrid under 150 random load scenarios. These 1-s data windows are trimmed into fifty 20-sample windows as shown in the right of Fig. 6. These 20 samples cover about 1.2 cycles of 60 Hz voltage signal. Thereafter, 55,770 graph data of 20-sample windows for the fault cases and 8,580 graph data of non-fault cases with random load changes are gathered as the train set. We also select 8,580 fault and 1,420 non-fault cases for the test set. Table I summarizes these configurations for fault cases and load changes data generation.

For the 123-node feeder system, similarly, we ran and captured totally 41,250 1-s window data of 11 fault types, 3 fault resistances under 50 random load scenarios at 25 buses (7,13, 18, 21, 25, 29, 35, 42, 47, 51, 53, 55, 57, 62, 65, 72, 80, 83, 86, 89, 93, 97, 99, 101). These 1-s data windows are also trimmed into fifty 20-sample windows and randomly selected for 41,250 fault samples. 10,000 load changes are generated to form the train set of the 123-node feeder with 33,000 fault samples. 10,000 mixed of 8,250 faults and 1,420 load-change samples are selected for the test set as shown in Table II.

### C. Training and Hyper-Parameter Tuning

The dataset is trained with Adam optimizer under the cross-entropy loss for binary classification in cases of fault event detection and cross-entropy loss for multi-class classification in cases of fault type and fault phase classification. To alleviate the overfitting problem, the random dropout of 10% is added in dense layers [51]. Batch size is a key hyper-parameter that decides the training time and model performance [52]. When increasing the batch size, we achieve a better approximation of gradient; however, the computational cost is significantly increased. In graph data, one graph already included the batch of all node data so that the appropriate batch size also depend on the size of the graph. Fig. 7 shows the relative time to complete 50 epochs and time to achieve 98% accuracy using the proposed R-GCN structure in binary fault/non-fault classification under the batch sizes of 50, 150, 250, 400. As can be seen, a good batch size is about 250. Notably, this hyper-parameter is relative since we trained on a personal computer with Intel Core i7-8700, 32 GHz, 32 GB RAM and NVIDIA GTX 1080 GPU. The machine learning framework is Pytorch with Pytorch-geometric library for graph learning [53].

The learning rate is also important to achieve high accuracy in a classification problem [54]. The learning rate used in the training process is started with 0.01 and then change to 0.001. Fig. 8 shows the training process when changing the learning rate from 0.01 to 0.001 after 120 epochs. The accuracy is saturated around 98.5% when the learning rate is kept at 0.01.

## IV. Comparative Numerical Results

The numerical results for the proposed fault diagnostics scheme using R-GCN are compared with popular neural networks structure such as ANN, LSTM, CNN and GCN. The details of these reference NN structures and the proposed R-GCN for fault event binary classification are described in

TABLE III. COMPARISONS OF NEURAL NETWORK STRUCTURES

| ANN | | LSTM | | CNN | | GCN | | R-GCN | |
|---|---|---|---|---|---|---|---|---|---|
| **Shared feature extration layers** | | | | | | | | | |
| Input | [780] | Input | [39×20] | Input | [39×20] | Input | [13×3×20] | Input | [13×3×20] |
| Dense | [512] | LSTM | [13×20] | CNN+pooling | 52×[20×10] | GCN | [13×24] | LSTM | [13×5×20] |
| Dense | [128] | LSTM | [3x20] | CNN+pooling | 26×[4×2] | GCN | [13×8] | GCN | [13×8] |
| Params: 469,664 | | Params: 4,000 | | Params: 3,994 | | Params: 3,344 | | Params: 2,694 | |
| **Fault event binary classification – Dense layers** | | | | | | | | | |
| Dense | [32] | Dense | [16] | Dense | [16] | Dense | [16] | Dense | [16] |
| Dense | [1] | Dense | [1] | Dense | [1] | Dense | [1] | Dense | [1] |
| **Fault location – Dense layers** | | | | | | | | | |
| Dense | [64] | Dense | [26] | Dense | [39] | Dense | [13×3] | Dense | [13×8] |
| Dense | [13] | Dense | [13] | Dense | [13] | Dense | [13] | Dense | [13] |
| **Fault type classification– Dense layers** | | | | | | | | | |
| Dense | [64] | Dense | [32] | Dense | [32] | Dense | [32] | Dense | [32] |
| Dense | [6] | Dense | [6] | Dense | [6] | Dense | [6] | Dense | [6] |
| **Fault phase classification– Dense layers** | | | | | | | | | |
| Dense | [64] | Dense | [32] | Dense | [32] | Dense | [32] | Dense | [32] |
| Dense | [3] | Dense | [3] | Dense | [3] | Dense | [3] | Dense | [3] |

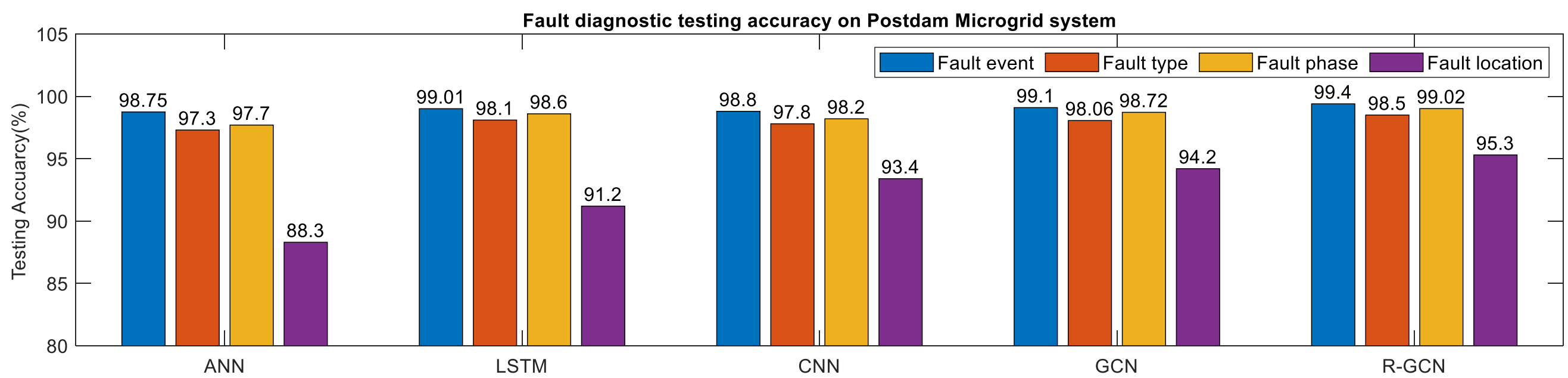

Fig. 10. Fault diagnostic accuracy of Potsdam Microgrid system using proposed R-GCN in comparison with ANN, LSTM, CNN and GCN structures.

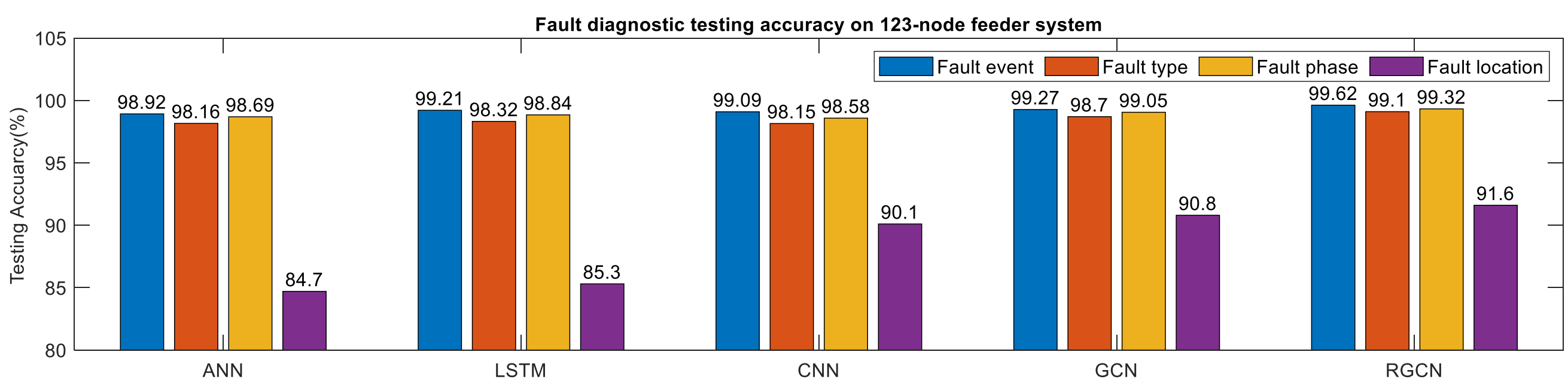

Fig. 11. Fault diagnostic testing accuracy of 123-node feeder system using proposed R-GCN in comparison with ANN, LSTM, CNN and GCN structures.

Table III, where the left columns show the operational layers, and the right columns show the sizes of the end-tensors. Reshaping and flattening operations are applied appropriately to condition the dimension compatibility between layers.

There is a shared feature extraction structures for all the tasks in fault diagnosis including fault event detection, fault phase, type classification, and fault location. First, the NN structures are trained with this feature extraction structure for the fault event binary classification with the intermediate dense layers as shown in Table III. Thereafter, these dense layers of binary classification are cut off and replaced by other dense layers for each task of fault location, fault type classification, and fault phase identification, respectively.

### *A. Comparison of Neural Networks Structures*

In comparison between the implemented NN structures i.e., ANN, LSTM, CNN, GCN, and the proposed R-GCN, first, the trainable parameters of ANN are significantly higher than other NN structures as shown in Table III. This is explained why the ANN can still achieve such that high accuracy. Fig. 9 shows the receiver operating characteristic (ROC) curves and the area under these curves (AUC) of the fault event binary classification models using the implemented NN structures. As can be seen, the proposed R-GCN structure has the highest

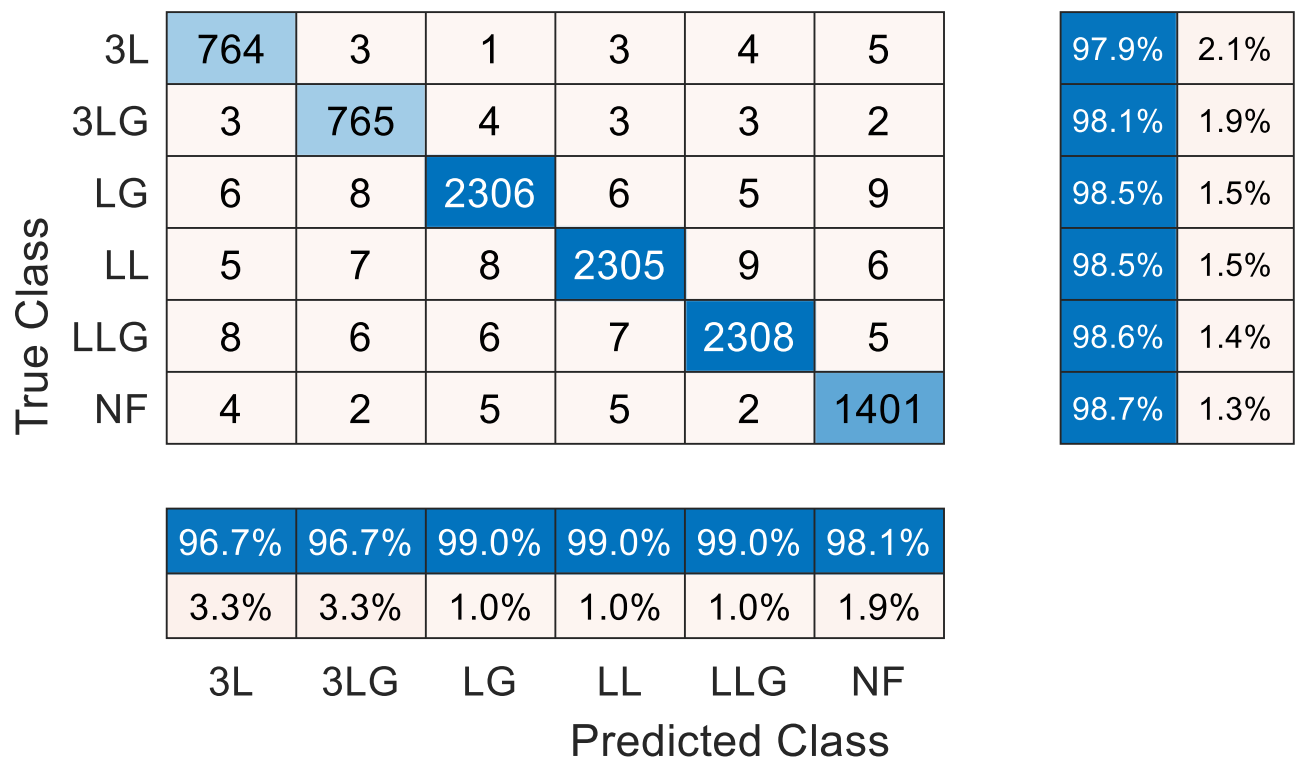

Fig. 12. Confusion matrix for fault type classification using R-GCN of Potsdam microgrid test set.

AUC at 0.998 compared to other structures, whereas the AUC of ANN is at 0.99757. Other LSTM, CNN, and GCN achieve around 0.998.

Figs. 10 and 11 show the testing accuracy in fault diagnostic tasks on the Potsdam microgrid and IEEE 123-node feeder datasets. In the figures, we compare the testing accuracy of fault event classification, fault type classification, fault phase classification, and fault location with different NN structures. From the numerical results, ANN can achieve 98.75% and 98.92% in both datasets while the proposed R-GCN can achieve 99.4% and 99.62% accuracy, respectively, in binary fault event detection. Other NN structures can also achieve around 99% to 99.2%. The CNN structure has a little bit less accuracy than LSTM and GCN. It may need a greater number of feature maps to achieve better performance.

In fault-type classification, the testing accuracy is slightly less due to multi-class classification. The proposed R-GCN achieve 98.5% and 99.1%. Here, we can see the testing accuracy in this IEEE 123-node feeder is higher than those of the Potsdam microgrid dataset despite the unbalance in the 123-node feeder. On one side, since the fault resistances are small, the unbalance does not affect the results. On the other side, there are 5 inverter-based generators in the Potsdam microgrid but only one source in the 123-node feeder. Once faults occur, the voltage drops in the 123-node feeder are more significant than those of the Potsdam microgrid. This explains why overall the testing accuracies in the 123-node feeder are higher than those of the Potsdam microgrid. Fig. 12 shows the detailed confusion matrix of fault type classification in the Potsdam microgrid, where the numbers in the diagonal show the samples predicted correctly and the numbers out of the diagonal indicate the samples mispredicted with other classes. The column on the right of the confusion matrix shows the percentages of sensitivity or recall or true positive rate (TPR) of each class, whereas the row under the confusion matrix shows the percentages of precision or positive predictive value (PPV) of each class.

Similarly, the fault phase classification achieves 99.02% and 99.32% accuracy with the proposed R-GCN. Other NN structures achieve around 98% while the GCN has 98.71% and 99.05% accuracy on two datasets, respectively. Figs. 13 and 14 show the detailed confusion matrix of fault phase identification for A/B/C and AB/BC/CA in the Potsdam microgrid, respectively.

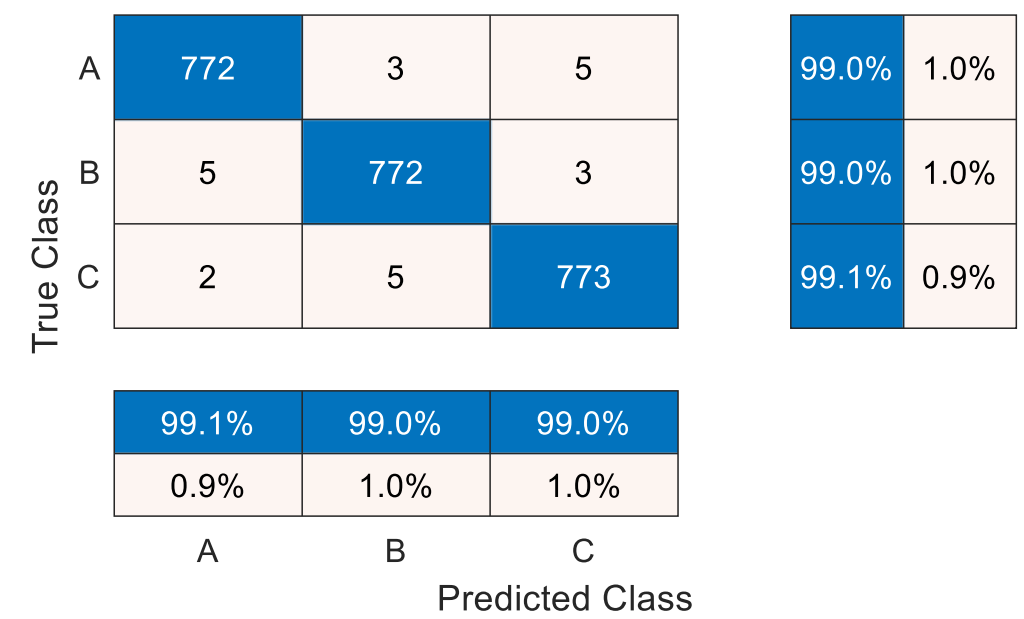

Fig. 13. Confusion matrix for fault phase A, B, and C classification using R-GCN of Potsdam microgrid test set.

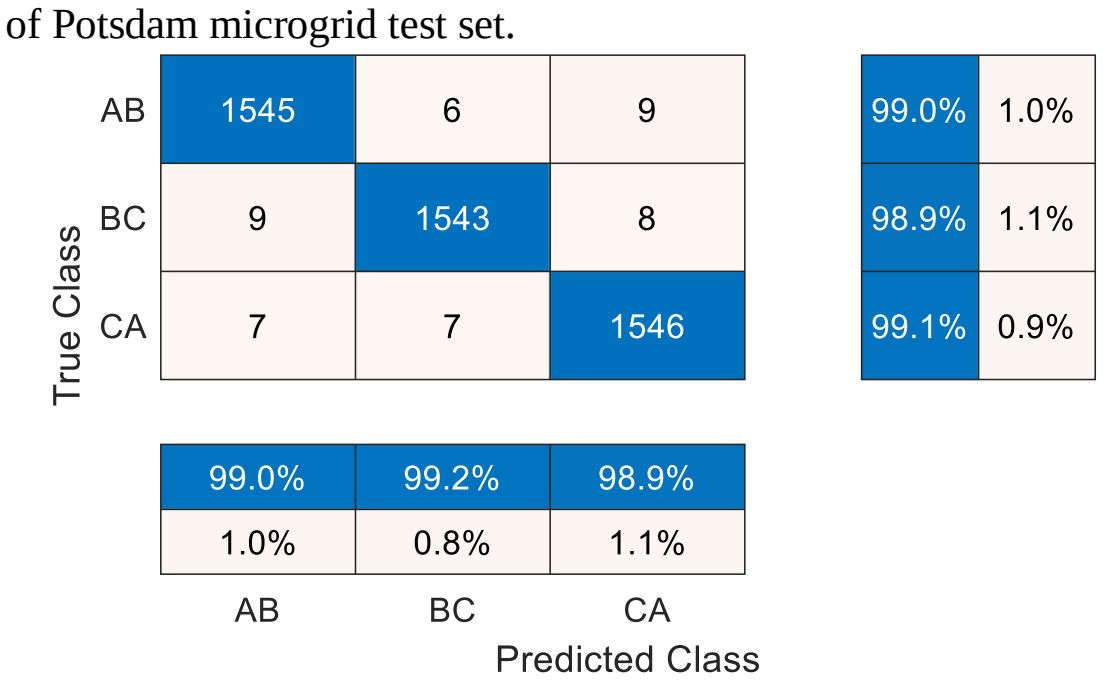

Fig. 14. Confusion matrix for fault phase AB, BC, and CA classification using R-GCN of Potsdam microgrid test set.

TABLE IV. COMPARISON WITH OTHER SCHEMES

| Scheme | Accuracy | | |
|---|---|---|---|
| | Event | Type | Phase |
| Decision tree [25] | 90.4% | 90.4% | 90.4% |
| K-nearest neighbors [25] | 95.63% | 95.63% | 95.63% |
| SVM [25] | 93.3% | 93.3% | 93.3% |
| Navie Bayers [25] | 94.24% | 94.24% | 94.24% |
| Wavelet-based GRU [29] | 99.31% | 97.6% | 97.92% |
| **GCN** | **99.1%** | **98.06%** | **98.72%** |
| **R-GCN** | **99.4%** | **98.5%** | **99.02%** |

TABLE V. IMPACT OF LESS VOLTAGE MEASUREMENTS

| NN | Accuracy % | | | | | |
|---|---|---|---|---|---|---|
| | Event | | Type | | Phase | |
| | PD | 123 | PD | 123 | PD | 123 |
| ANN | 98.12 | 98.34 | 96.93 | 97.65 | 97.15 | 98.1 |
| LSTM | 98.63 | 98.84 | 97.76 | 97.88 | 98.08 | 98.24 |
| CNN | 98.01 | 98.75 | 97.36 | 97.93 | 97.6 | 98.02 |
| GCN | 98.58 | 98.91 | 97.49 | 98.31 | 98.18 | 98.64 |
| R-GCN | 99.03 | 99.17 | 97.92 | 98.7 | 98.8 | 98.85 |

TABLE VI. IMPACT OF MEASUREMENT NOISES

| Noises (SNR) | Accuracy % of R-GCN | | | | | |
|---|---|---|---|---|---|---|
| | Event | | Type | | Phase | |
| | PD | 123 | PD | 123 | PD | 123 |
| No noises | 99.4 | 99.62 | 98.5 | 99.1 | 99.02 | 98.32 |
| 30 dB | 99.32 | 99.57 | 98.48 | 99.14 | 98.96 | 98.28 |
| 25 dB | 99.03 | 99.16 | 98.14 | 98.76 | 98.61 | 97.96 |
| 20 dB | 98.31 | 98.53 | 97.49 | 98.27 | 92.7 | 91.27 |

The accuracies in fault location are less than those of other classifiers. The proposed R-GCN can achieve 95.5% and 91.6% accuracies on the Potsdam microgrid and 123-node feeder, respectively. They are more than 7% compared to those of ANN structure. The fault location accuracies in Potsdam microgrid are much larger than the 123-node feeder due to its smaller size and the graph structure considering all buses. Notably, herein, we only simulate and detect faults that occur on the main buses of the target systems. Faults that occurred in between the connecting lines are considered to belong to the nearest bus.

The accuracy performances of GCN and proposed R-GCN in this paper are compared with existing schemes in Table IV. As can be seen, the proposed scheme can compete with state-of-the-art schemes, especially the wavelet-based deep-NN using GRU [55]. Notably, herein, we consider that voltage measurement while other schemes have branch currents as inputs. Therefore, the proposed scheme should be superior to the existing methods.

### *B. Impact of Different Measurement Conditions*

To investigate the impact of fewer voltage measurements, we perform the fault diagnostic scheme again with only 3 voltage measurements from buses 1, 5, and 9 in the Potsdam microgrid. In the IEEE 123-node feeder, we drop voltage measurements on buses 1, 25, 47, 63, 91, and 105 to zero, while keeping the voltage measurements on buses 13, 18, 54, 76, 97, and 300. The accuracies of both the Potsdam microgrid (denoted as PD) and IEEE 123-node feeder (denoted as 123) under this fewer voltage measurements are shown in Table V. The accuracies in all fault event detection, fault phase/type classification is less than around 0.4% to 0.8% compared to those with all measured voltages. Notably, here, the remained voltage measurements are still scattered all the investigated system. In future works, we may consider the loss of all voltage measurements in a certain area of the systems.

To investigate the impact of measurement noises, we add the noises with zero mean, Gaussian distribution, and signal-to-noise ratios of 30 dB (3.2%), 25 dB (5.6%), and 20 dB (10%) to the voltage measurement before training and testing. The accuracies of both the Potsdam microgrid (denoted as PD) and IEEE 123-node feeder (denoted as 123) under these additional noises are shown in Table VI. As can be seen, the influences of small noises are insignificant since the accuracies only drop about 0.1% and 0.3% under 30 dB and 25 dB noises respectively. However, under 20 dB of noise, the accuracies reduce by about 1%. It is projected to be worse under more noise.

## V. Conclusion

In this paper, we proposed an R-GNN structure of fault diagnostic including fault detection, fault type, phase identification, and fault location utilized voltage measurements in power distribution systems. The EMT datasets of the Potsdam microgrid and IEEE 123-node feeder of under faults and load changes are created in Opal-RT real-time simulator and can be utilized for further research. The transfer learning and fine-tuning technique are applied to reduce the training effort. The performance of the proposed R-GNN structure is compared with the benchmarking NN structures i.e. ANN, LSTM, CNN, and GCN. The numerical results show that the proposed R-GNN achieved state-of-the-art accuracies compared to existing NN structures and other schemes. The impact of fewer voltage measurements and measurement noises is investigated. The numerical results demonstrate the superiority of the proposed fault diagnostic scheme using R-GCN.

## Acknowledgement

We would like to thank Shuvangkar Chandra Das and Quang-Ha Ngo for helping the real-time simulation on Opal-RT, data collection and discussion.

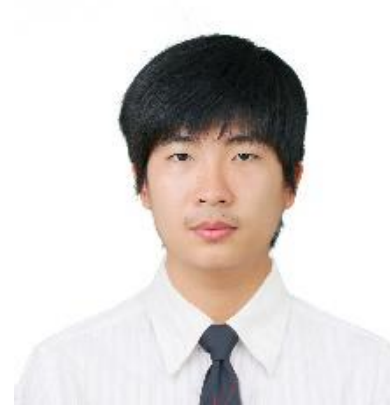

**BANG L.H. NGUYEN** (St.M) received the B.Eng. degree and M.Eng. degree in Electrical and Electronics Engineering from the VNU HCMC – University of Technology, Vietnam, in 2010 and 2013, respectively. In 2015, he was with the Eastern International University, Binh Duong, Vietnam as a lecturer. From 2016 to 2018, he was a research assistant in the power electronics and energy conversion lab (PEEC), Kyungpook National University, Korea. He is currently working toward the Ph.D. degree in Smart Power Systems and Controls Lab, Clarkson University, Potsdam, New York, USA. From Jun. 2021, he is with National Renewable Energy Laboratory as a graduate student researcher. Bang Nguyen was a recipient of the IEEE Industrial Electronics Society Best IEEE Industrial Electronics Magazine in 2021. His research expertise include design and control of power electronics system, design and control of power and energy systems included PHIL and CHIL.

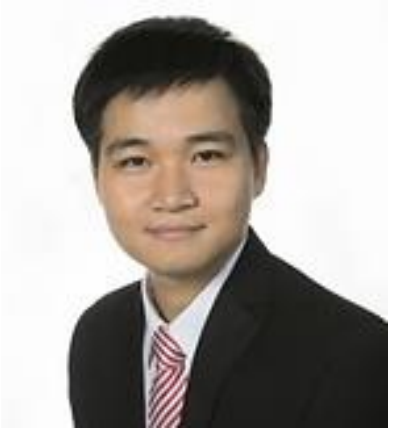

**TUYEN (TONY) VU** (M) received the B.S. degree in electrical engineering from the Hanoi University of Science Technology, Hanoi, Vietnam, in 2012, and the Ph.D. degree in electrical engineering from Florida State University, Tallahassee, FL, USA, in 2016. From 2016 to 2017, he was a Postdoctoral Research Associate with the Center for Advanced Power Systems, Florida State University, where he was a Research Faculty from 2017 to 2018. Since July 2018, he has been an Assistant Professor with Clarkson University, Potsdam, NY, USA. Tony was the lead author the IEEE Industrial Electronics Society - Best IEEE Industrial Electronics Magazine in 2021. Tony was a Guest Editor for IEEE Transactions on Industrial Informatics. His research interests include smart grid; power system dynamics, stability, and control; energy management and optimization; power systems cybersecurity, and integration of distributed energy resources into power system.

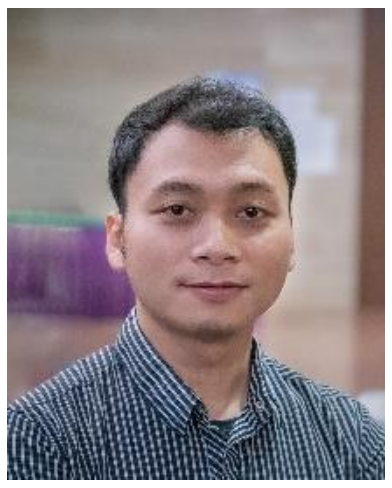

**THAI-THANH NGUYEN** (M) received the B.S. degree in electrical engineering from the Hanoi University of Science and Technology, Vietnam, in 2013, and the Ph.D. degree in electrical engineering from Incheon National University, South Korea, in 2019. From 2019 to 2022, he was a Postdoctoral Researcher and a Research Professor with Incheon National University, South Korea, and a Research Associate at Clarkson University, USA. Since April 2022, he has been an Engineer Scientist with Advanced Grid Innovation Laboratory for Energy (AGILe), New York Power Authority, USA. His research interests include power system modeling and control, power converter control, and application of power electronics to power systems.

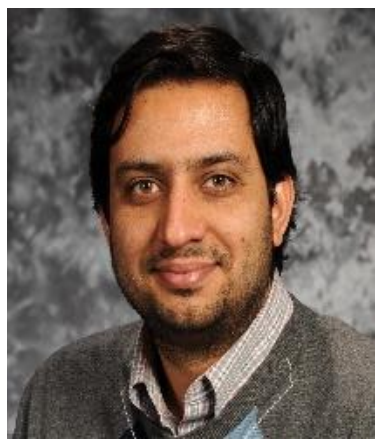

**MAYANK PANWAR** (M'11) is a Research Engineer in the Power Systems Engineering Center at National Renewable Energy Laboratory in Golden, CO. He received his Ph.D. and M.S. degrees in Electrical Engineering from Colorado State University in Ft. Collins, CO in 2017 and 2012, respectively. He has a B.Tech. degree in Electronics and Instrumentation engineering from A.P.J. Abdul Kalam Technical University (formerly U.P. Technical University), India in 2007. He was a Postdoctoral Researcher from Feb. 2017 to Nov. 2017, Research Scientist from Nov. 2017 to Dec. 2019 in the Power and Energy Systems department at Idaho National Laboratory, Idaho Falls, ID, U.S.A. From 2007-2011, he was with NTPC Limited, India where he worked as a Control and Instrumentation Engineer. He has worked on several DOE-funded projects including GMLC RADIANCE in Alaska for resilient distribution systems. His research interests include microgrids, real-time simulations, hardware-in-the-loop testing, hydropower, co-simulation of electrical-mechanical-thermal systems, and machine learning applications in power systems.

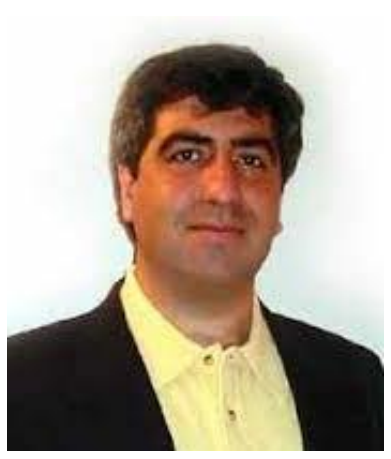

**ROB HOVSAPIAN** (SM) received the M.S. degree in control and the Ph.D. degree in energy systems from the Mechanical Engineering Department, Florida State University, Tallahassee, FL, USA, in 1988 and 2009, respectively. He has spent more than 20 years working with Idaho National Laboratory, General Dynamics, TRW, and Northrop Grumman, as a Research Faculty with the Mechanical Engineering Department and as a Program Manager with the Office of the Naval Research, Center for Advanced Power Systems for the Electrical Ship Research and Development Consortium, Florida State University. He is currently working with the National Renewable Energy Laboratory, Golden, CO, USA, as a Research Advisor. He has a number of publications in the field of energy systems, thermodynamics optimization, thermal modeling, wind energy, and controls.